\documentclass[onecolumn, draftclsnofoot, 12pt]{IEEEtran}
\usepackage{comment}
\usepackage{cite}
\usepackage{amsmath,amssymb,amsfonts}
\usepackage{algorithmic}
\usepackage{graphicx}
\usepackage{textcomp}
\usepackage{xcolor}
\usepackage{comment}
\usepackage{multirow}
\usepackage{url}
\usepackage{makecell}
\usepackage{soul}

\ifCLASSINFOpdf
\else

\fi

\begin{document}

\title{Development of a Compact High-Voltage Functional Electrical Stimulation Device}

\author{Haiduo Wang, Ruizhe Tang,
        and~Xilin~Liu\\
        Department of Electrical and Computer Engineering, University of Toronto
}

\maketitle

\begin{abstract}
This report details the design and development of a compact high-voltage functional electrical stimulation (FES) device. Unlike conventional FES systems, the proposed design prioritizes user comfort by leveraging rapid switching times to effectively activate muscles while minimizing stimulation of pain receptors. The device is equipped with a high compliance voltage of up to 135 V, enabling its use across various muscle groups and accommodating users with differing skin conductance levels. At the core of the system is a switched capacitor (SC) stimulator, which facilitates fast switching, and a flyback converter that steps up the battery voltage. The design exclusively utilizes off-the-shelf components, significantly reducing prototyping costs. Initially, a large testing board was developed to evaluate the performance of the SC stimulator in comparison with conventional H-bridge-based current-regulated stimulators. Subsequently, a compact, three-board PCB design was created to enable battery-powered operation, making it suitable for wearable applications. The proposed design methodology is broadly applicable to a wide range of neural rehabilitation applications.
\end{abstract}


\IEEEpeerreviewmaketitle

\section{Introduction}

Functional Electrical Stimulation (FES) is an advanced therapeutic technique that employs carefully controlled electrical pulses to stimulate excitable tissues, such as peripheral nerves and muscles, with the aim of restoring lost motor functions \cite{popovic2014advances}. By bypassing damaged neural pathways, FES directly induces muscle contractions, effectively bridging the gap between the central nervous system and the affected musculature. This stimulation not only generates immediate muscle movements but also fosters long-term neuroplastic changes within the central nervous system. These changes are crucial as they facilitate the relearning of motor control, enabling patients to regain voluntary movement over time \cite{peckham2005functional}.

The clinical applications of FES are broad, ranging from rehabilitation following spinal cord injuries and strokes to assisting individuals with neuromuscular diseases such as multiple sclerosis or cerebral palsy. FES has demonstrated significant benefits in reducing muscle atrophy, improving circulation, and maintaining joint integrity in patients with limited mobility. Moreover, its use in promoting neuroplasticity highlights its role not just as a passive therapeutic tool, but as an active participant in reshaping and retraining the brain and spinal cord to restore functional independence.

Given its potential, FES has garnered considerable attention in the medical community, paving the way for innovative solutions to address motor deficits. Modern advancements have expanded its applications, such as in wearable devices for daily support, and its integration with brain-computer interfaces for enhanced precision. By improving motor functions and enhancing the quality of life for individuals with disabilities, FES continues to redefine the scope of neurorehabilitation and functional recovery~\cite{lynch2008functional}.

However, a significant limitation to the widespread adoption of FES is the pain associated with its use. This discomfort arises because the electrical stimulation not only activates the target muscles but also stimulates pain receptors in the process, leading to varying levels of patient intolerance \cite{Huerta2012}. This issue is further compounded by the high compliance voltage required for effective stimulation, which is heavily dependent on the user’s skin conductance. For some individuals, particularly those with higher skin impedance, the compliance voltage can exceed 100 V during therapeutic sessions.

This high voltage introduces multiple technical challenges in the design and development of FES devices. First, it surpasses the breakdown voltage of standard CMOS circuit components, necessitating specialized high-voltage (HV) circuitry. Incorporating such HV components not only increases the complexity of the system but also adds to the cost and energy consumption, limiting the feasibility of creating compact and cost-efficient devices. Furthermore, the challenges of managing heat dissipation and ensuring user safety at such high operating voltages further complicate the design process.

As a result, most commercially available FES devices tend to be bulky and non-portable, often designed as rack-mounted systems intended for use in clinical or research settings. These devices are generally unsuitable for wearable applications, restricting their accessibility and practicality for daily use by patients. The lack of portability is particularly problematic, as consistent and frequent use of FES is often necessary to achieve optimal therapeutic outcomes, such as muscle reconditioning and neuroplasticity.

Innovations in miniaturized, high-voltage circuit design, such as switched capacitor stimulators and flyback converters, offer potential solutions to these challenges. By efficiently stepping up battery voltages and integrating advanced safety features, these technologies could enable the development of smaller, more user-friendly FES devices suitable for wearable applications. Nevertheless, achieving a balance between miniaturization, safety, and effectiveness~\cite{van2016system} remains a key hurdle in advancing FES technology for broader clinical and at-home use.

This study introduces a wearable high-voltage (HV) FES device designed to overcome the two primary challenges associated with conventional FES systems. First, the switching time of the proposed device is reduced to less than 20~ns through the implementation of a SC circuit, as opposed to traditional current-mode stimulation techniques \cite{vidal2010towards,liu2020fully}. This rapid switching enhances the device’s efficiency and performance while minimizing the stimulation of pain receptors. Second, the device achieves a high compliance voltage through the use of a flyback regulator, which is optimized to prevent excessive power consumption and heat generation.

This dual approach not only addresses the pain-related challenges of FES but also significantly reduces the size and complexity of the circuitry. The integration of these design elements is critical to minimizing the overall footprint of the system, ensuring its viability for wearable applications. By combining advanced circuit design with efficient power management, the proposed FES device offers a compact and portable solution suitable for therapeutic use in daily settings~\cite{blade2023semg}.

\section{Switched-Capacitor FES Design}

\subsection{Design Requirements}

\begin{figure}[!ht]
    \centering
    \includegraphics[width=.8\linewidth]{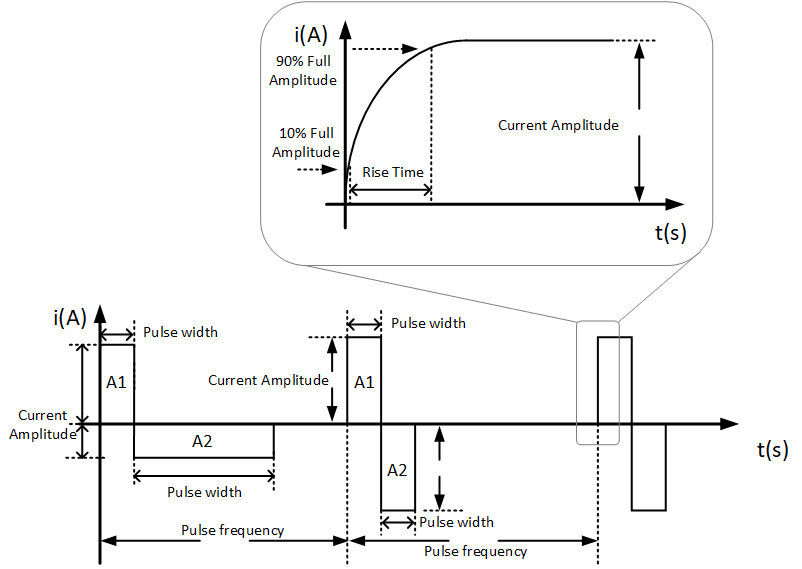}
    \caption{An illustration of the proposed multi-modal sensing system for at-home rehabilitation exercise monitoring}
    \label{fig:intro}
\end{figure}

Biphasic stimulation, utilizing either symmetrical or asymmetrical rectangular pulses, is the most commonly employed method in FES. The waveform of a typical biphasic stimulation is depicted in Fig.~\ref{fig:intro}. Key adjustable parameters in FES therapy include stimulation intensity, which is typically quantified by stimulation current or charge density, stimulation frequency, and the duration of the stimulation pulse \cite{liu2015pennbmbi}. These parameters are critical in tailoring the therapeutic outcomes for individual patients, ensuring both efficacy and safety.

One parameter that is often underestimated in FES design is the switching time of the stimulation pulses, particularly the rise time of the stimulation phase. Rise time refers to the time required for an electrical pulse to transition from its resting potential to its peak voltage. This characteristic has significant implications for both muscle activation selectivity and patient comfort \cite{ibitoye2016strategies}. Research indicates that shorter rise times, typically under 100~ns, enable faster muscle contractions and can reduce the current amplitude required for effective muscle activation. In contrast, prolonged rise times may lead to increased muscle fatigue and patient discomfort during therapy. Thus, optimizing rise time is a critical design consideration for enhancing patient comfort, improving therapeutic outcomes, and increasing the energy efficiency of FES devices.

Additionally, achieving high stimulation voltages is paramount in FES systems to activate larger or deeper muscles and to overcome the impedance presented by skin and underlying tissues. High-voltage stimulation ensures that sufficient current reaches target muscles, even in deeper layers, by effectively penetrating through skin resistance and other tissue barriers. This capability is essential for patients requiring the activation of less accessible muscle groups or those with higher skin resistance. Therefore, the integration of high-voltage capabilities, combined with optimal rise time control, is fundamental in advancing FES technology to deliver precise, comfortable, and energy-efficient therapy.

\subsection{Operational Principle}

\begin{figure}[!ht]
    \centering
    \includegraphics[width=.8\linewidth]{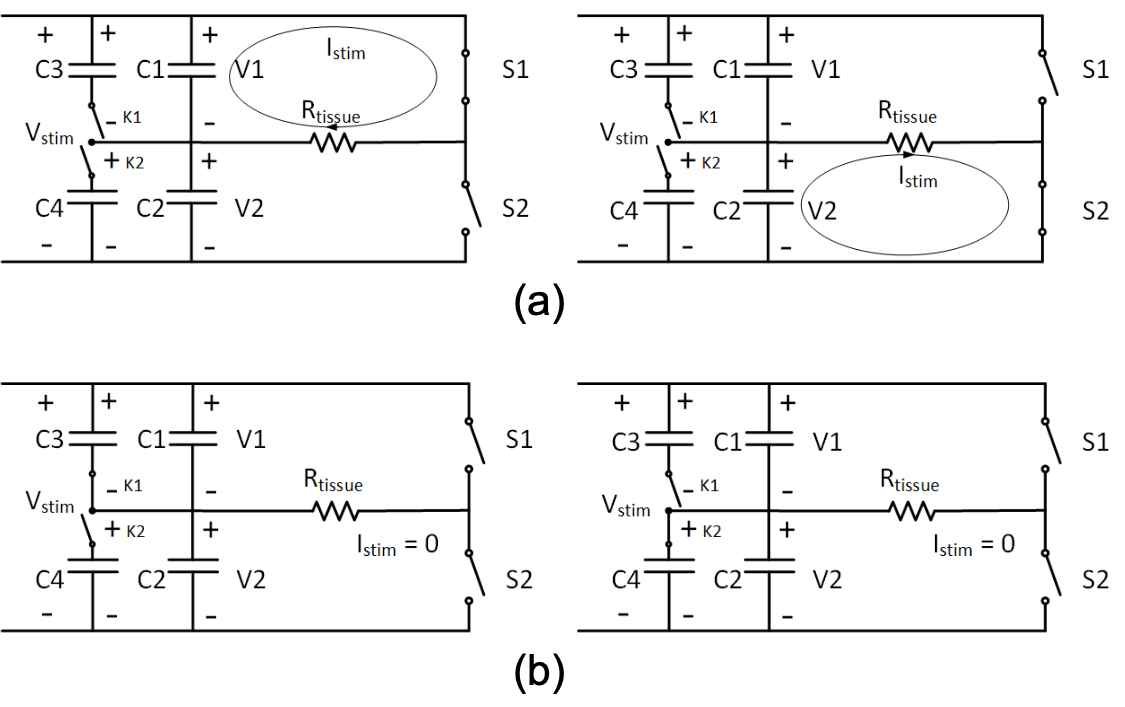}
    \caption{(a) Illustration of the operation of the SC stimulator. (b) Generation of asymmetric pulses.}
    \label{fig:scheme}
\end{figure}

To achieve both high voltage capability and rapid switching times, a SC circuit-based stimulator is utilized in this work. The operational principle of the SC-based stimulator is illustrated in Fig.~\ref{fig:scheme}(a). During the stimulation phase, switch S1 is closed (ON) while switch S2 is open (OFF), resulting in a current flow through the tissue load. Conversely, during the reversal phase, S1 is open and S2 is closed, causing the current to flow in the opposite direction through the load. This bidirectional current flow ensures effective stimulation while maintaining operational flexibility.

The design incorporates four capacitors (C1–C4) of equal capacitance, configured to create a dynamically adjustable capacitive divider. As depicted in Fig.~\ref{fig:scheme}(b), C1 and C2 are permanently connected, while C3 and C4 can be selectively connected via solid-state relays (K1 and K2). This configuration enables the generation of both symmetric and asymmetric output pulses.

When both relays (K1 and K2) are off, the SC circuit operates in a symmetric mode, producing equal voltages $V_1$ and $V_2$ across the load. In contrast, activating K1 while keeping K2 off causes the circuit to operate in an asymmetric mode, yielding a voltage ratio of $V_1$:$V_2$ = 1:2. This results in a stimulation current amplitude ratio of 1:2 and a corresponding pulse width ratio of 2:1. Conversely, when K1 is off and K2 is on, the system generates a current amplitude ratio of 2:1 and a pulse width ratio of 1:2.

This flexible configuration allows the SC stimulator to adapt to varying therapeutic requirements by precisely controlling current amplitude and pulse width ratios. Such adaptability is essential for tailoring stimulation patterns to optimize therapeutic efficacy while minimizing discomfort, making the design highly suited for diverse FES applications.

\section{Circuits and System Implementation}

\subsection{System Overview}

A high-level block diagram of the proposed FES device is presented in Fig.~\ref{fig:block_diagram}. The system architecture is composed of three main modules: the power stage, the stimulation output stage, and the control module. Each module is designed to work in harmony to deliver efficient, high-voltage stimulation with precise control, meeting the requirements of wearable FES devices. The components and functionality of each module are described as follows:

\begin{figure}[!ht]
    \centering
    \includegraphics[width=1\linewidth]{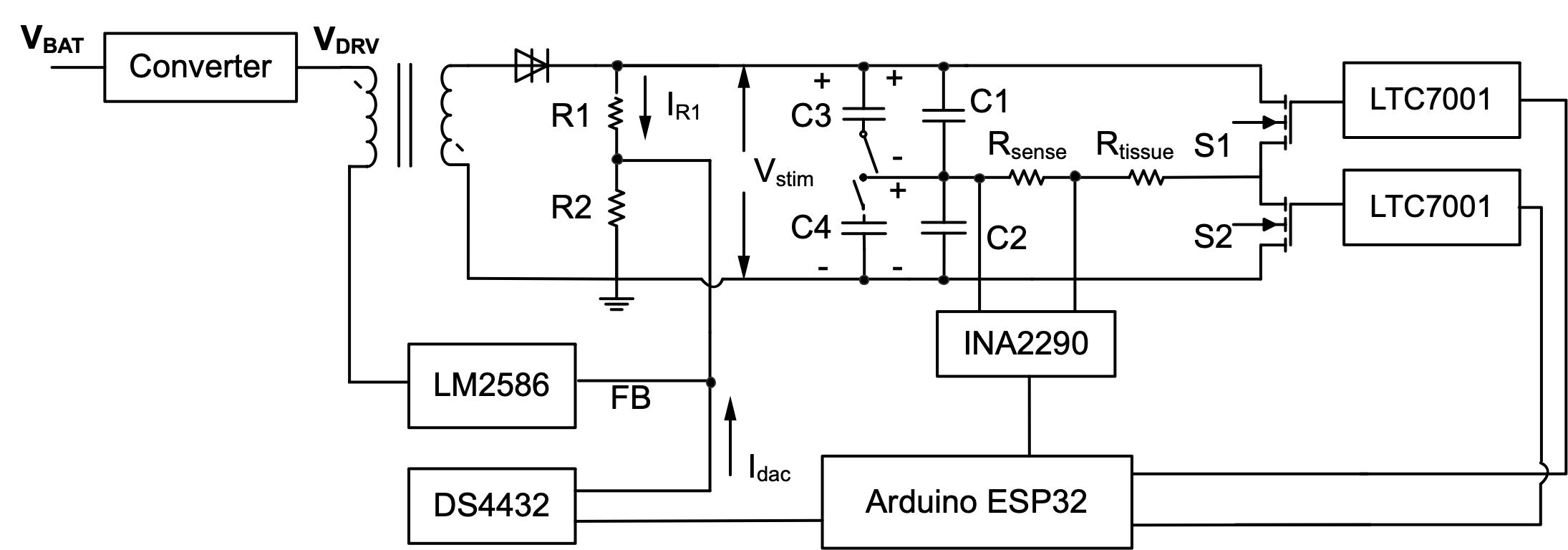}
    \caption{The overall block diagram of the proposed HV FES device.}
    \label{fig:block_diagram}
\end{figure}

\subsubsection{Power Stage}
The power stage consists of two converter blocks that ensure the device can operate reliably with minimal power loss. The first block is a low-voltage regulator that stabilizes the input battery voltage at 12V. This regulated voltage powers the low-voltage components, including the control module and auxiliary circuits. The second block is a high-voltage flyback converter, equipped with a resistive feedback divider, which steps up the 12V to the desired compliance voltage. This flyback converter can generate an output voltage of up to 135~V, providing the necessary power to stimulate deeper or larger muscle groups while overcoming skin and tissue impedance. This dual-stage approach ensures both low-power operation and high-voltage output, a critical requirement for compact and wearable applications.

\subsubsection{Stimulation Output Stage}
The stimulation output stage incorporates the SC-based stimulator circuit, which is central to the device’s ability to produce high-performance stimulation pulses. The SC stimulator is regulated by the control module and is designed to generate pulses with a high slew rate, ensuring rapid muscle activation while minimizing patient discomfort. This stage supports both symmetric and asymmetric pulse configurations, allowing for customizable therapy tailored to individual needs. The high slew rate and precise output control contribute to the device’s efficiency, reducing both power consumption and heat generation.

\subsubsection{Control Module}
The control module is powered by an Arduino Nano ESP32 microcontroller, which acts as the brain of the system. It is responsible for configuring stimulation parameters such as pulse width, frequency, and amplitude. The microcontroller orchestrates the operation of the SC stimulator by generating precise timing signals and controlling the solid-state relays for asymmetric pulse generation. Additionally, it interfaces with a current sensor to monitor the stimulation current in real time, dynamically adjusting the output voltage to maintain stable and effective operation under varying load conditions. The use of the Arduino Nano ESP32 enables a compact and flexible design with ample processing power for future enhancements, such as wireless control or integration with wearable monitoring systems.

This modular design ensures scalability and adaptability, enabling the proposed FES device to meet a wide range of therapeutic requirements. The following subsections delve deeper into the design and operation of the flyback regulator and the SC stimulator, detailing their contributions to the overall performance of the system.

\subsection{Flyback Converter with Resistive Feedback Divider}

\begin{figure}[!ht]
    \centering
    \includegraphics[width=.7\linewidth]{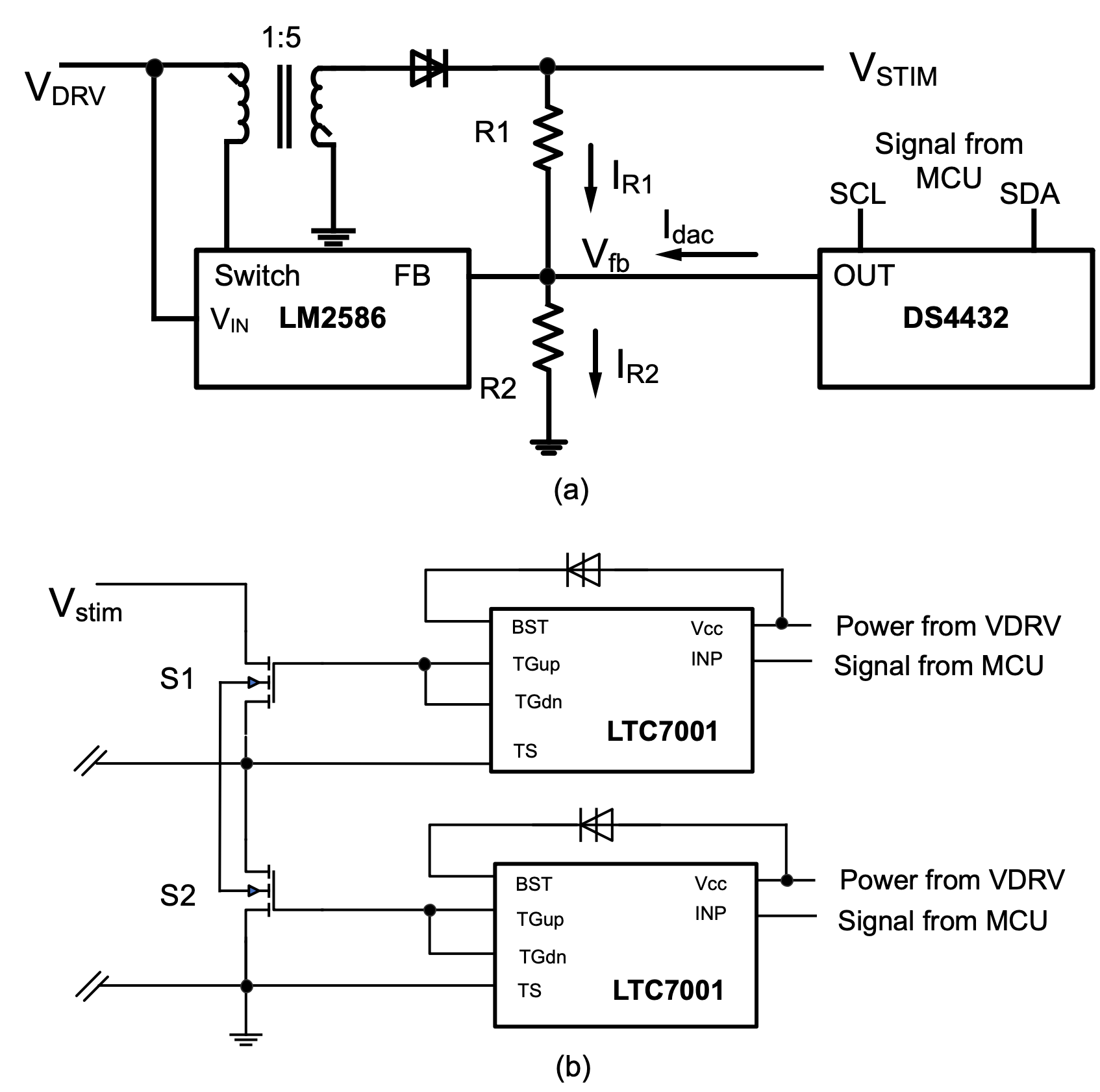}
    \caption{(a) The detailed diagram of the flyback power converter. (b) The detailed diagram of the HV switches of the SC circuits with the gate drivers.}
    \label{fig:ckt}
\end{figure}

The implementation details of the flyback converter, utilizing the LM2586 regulator from Texas Instruments, are illustrated in Fig.~\ref{fig:ckt}(a). This flyback converter plays a critical role in generating the high compliance voltage required for effective FES while maintaining system efficiency and stability. During operation, the LM2586 employs an internal switch that alternates between closed and open states. When the switch is closed, current flows through the primary winding of the transformer, storing energy in its magnetic field. When the switch opens, this stored energy is transferred to the secondary winding, generating the desired output voltage. The design leverages this energy transfer mechanism to achieve a compact and efficient voltage boost necessary for high-voltage applications.

The output voltage of the converter is determined by the ratio of the high-side resistor $R_1$ to the low-side resistor $R_2$, which forms part of the feedback loop. The feedback voltage ($V_{fb}$) at the feedback node is maintained at a constant value to match the reference voltage ($V_{ref}$), ensuring stable operation and precise controlof the output voltage ($V_{out}$).

To enable dynamic adjustment of the stimulation voltage, a current mode digital-to-analog converter (IDAC), specifically the DS4432 from Analog Devices Inc., is integrated into the feedback path. This IDAC provides fine control by sourcing or sinking current, allowing for real-time modulation of $V_{out}$.
\begin{itemize}
    \item When the IDAC sources current into the feedback loop, the feedback voltage ($V_{fb}$) increases. To maintain the constant feedback condition, the current through $R_1$ ($I_{R1}$) decreases, resulting in a reduction in the output voltage ($V_{out}$).
    \item Conversely, when the IDAC sinks current from the feedback loop, $I_{R1}$ increases to stabilize $V_{fb}$, thereby increasing the output voltage ($V_{out}$).
\end{itemize}

This dynamic voltage adjustment mechanism ensures that the output voltage can be precisely tailored to meet varying therapeutic needs, such as stimulating deeper muscle groups or adapting to differences in user skin impedance. Moreover, the use of a programmable IDAC in conjunction with the flyback converter introduces flexibility in the system, allowing it to adapt to real-time feedback from other system components, such as the current sensor or microcontroller.

The design not only optimizes the efficiency of the power stage but also contributes to the overall compactness of the device, making it suitable for wearable applications. The capability to fine-tune $V_{out}$ dynamically is particularly advantageous in enhancing the comfort and effectiveness of FES therapy, ensuring the device can accommodate a wide range of users and therapeutic scenarios.

The flyback converter with a resistive feedback divider serves as the foundational power stage architecture of the stimulator. This design incorporates a programmable current-output digital-to-analog converter (IDAC) connected to the feedback path, enabling dynamic adjustment of the output stimulation voltage. This adaptability is critical for tailoring the voltage output to accommodate a variety of therapeutic requirements, ensuring optimal performance across different use cases. At the core of this architecture is the LM2586 flyback converter, paired with a DS4432 7-bit dual-channel IDAC, forming a flexible and efficient power source. The system operates with a 12 V input voltage and utilizes a transformer with a 1:5 turns ratio, enabling the flyback converter to deliver an output voltage range of up to 120 V. The feedback voltage at the feedback node is maintained at a constant reference value of 1.23 V, ensuring stability and precision in voltage regulation.

The DS4432 IDAC modulates the output voltage dynamically. Specifically, when the IDAC sources 200~$\mu$A of current into the feedback path, the output voltage is reduced to as low as 3.5 V. Conversely, when the IDAC sinks 200~$\mu$A, the output voltage can reach a maximum of 120 V. With its 7-bit resolution, the IDAC offers 255 discrete voltage settings, providing a fine resolution of 0.457 V per step. This high level of granularity enables precise control over the stimulation voltage, ensuring that the device can be customized to the specific needs of individual users and therapeutic protocols. This programmable design offers several advantages for FES applications. First, the wide voltage range supports both superficial and deep muscle stimulation, making it suitable for a diverse population of users with varying tissue impedance. Second, the fine resolution ensures that incremental adjustments can be made, improving both the comfort and effectiveness of the therapy. Finally, the integration of the IDAC into the feedback loop streamlines the system’s complexity, reducing the need for additional hardware while maintaining a compact form factor.

\subsection{HV Switches and Gate Driver}
Achieving rapid switching times is a critical requirement for the performance and efficiency of the proposed system. This is particularly essential in FES applications, where precise and high-speed pulse generation directly impacts muscle activation and user comfort. In the proposed design, the combination of a high-side gate driver with an N-channel MOSFET provides a robust and efficient solution for achieving the desired fast switching performance.

The high-side gate driver (LTC7001, Analog Devices Inc.) is integral to this implementation. It employs an internal charge pump that enhances the gate voltage of the external MOSFET, enabling rapid and efficient transitions between the operating phases of the SC circuit. This internal charge pump allows the gate driver to operate effectively at high voltages, ensuring reliable and fast switching in demanding environments.

The LTC7001 is designed with a low pull-up resistance of 2.2~$\Omega$ and an even lower pull-down resistance of 1~$\Omega$, which contributes significantly to its rapid response times. These low resistances allow the driver to charge and discharge the gate of the MOSFET quickly, resulting in swift turn-ON and turn-OFF times measured in just a few nanoseconds. These characteristics are essential for maintaining the high slew rates required in the SC circuit, which, in turn, support the rapid rise and fall times of the stimulation pulses.

Extensive testing was conducted to evaluate the performance of various gate driver solutions. Among the options considered, the LTC7001 demonstrated superior reliability and speed, making it the most optimized choice for this design. Its performance in high-side switching applications ensures stable operation, even under the high compliance voltages necessary for FES. Additionally, its compact form factor and integration with external MOSFETs provide flexibility and scalability for future system enhancements.

The adoption of the LTC7001 not only enhances the speed and reliability of the switching mechanism but also contributes to the overall energy efficiency of the system. Faster switching minimizes power losses in the MOSFET during transitions, which is particularly important in battery-powered wearable devices. Furthermore, its robustness ensures consistent performance over extended periods, making it an ideal choice for medical applications where reliability and precision are paramount.

\section{Experimental Results}

\begin{figure}[!ht]
    \centering
    \includegraphics[width=.8\linewidth]{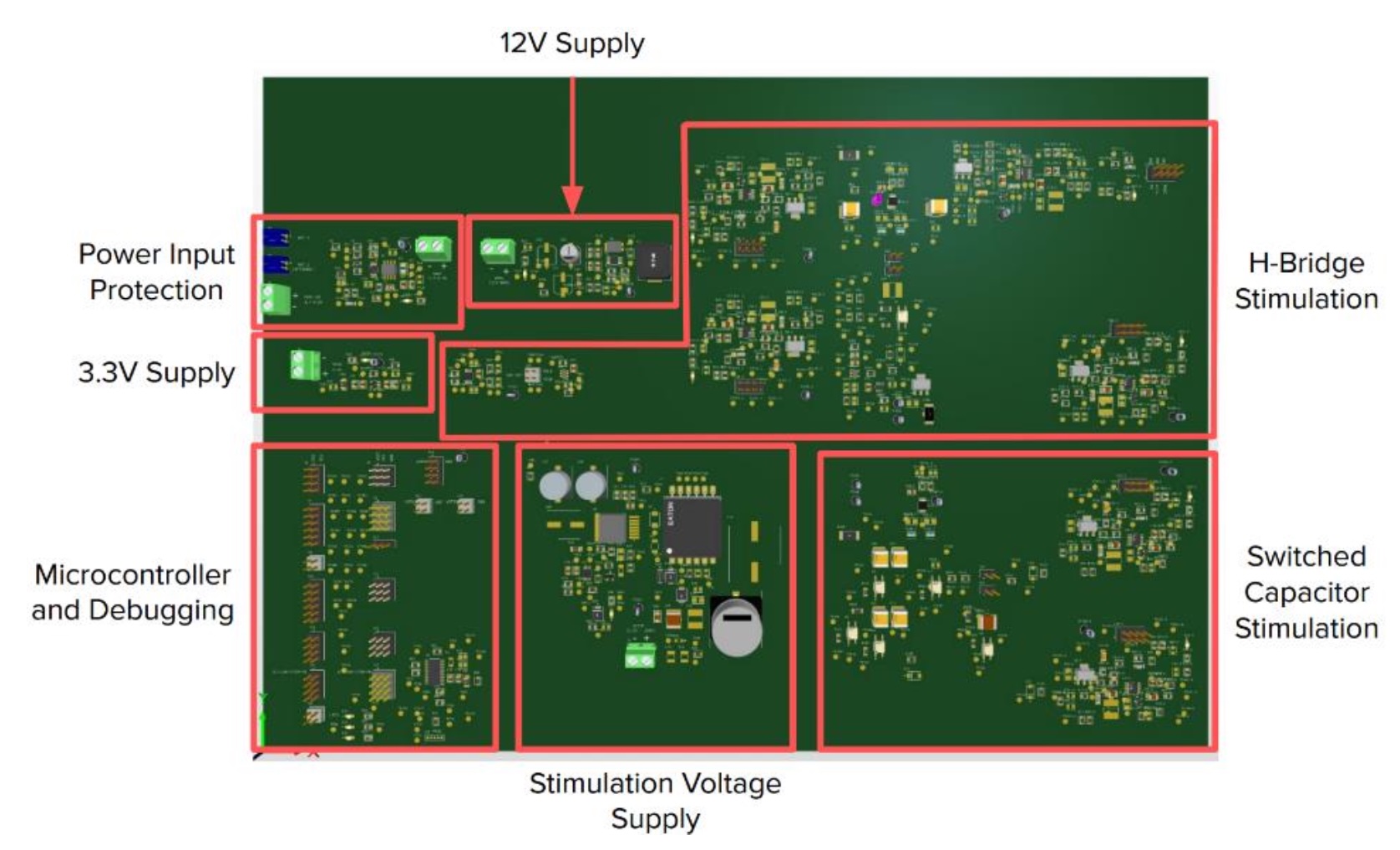}
    \caption{PCB layout of the large test board with key functional blocks highlighted. }
    \label{fig:PCB}
\end{figure}

\begin{figure}[!ht]
    \centering
    \includegraphics[width=.8\linewidth]{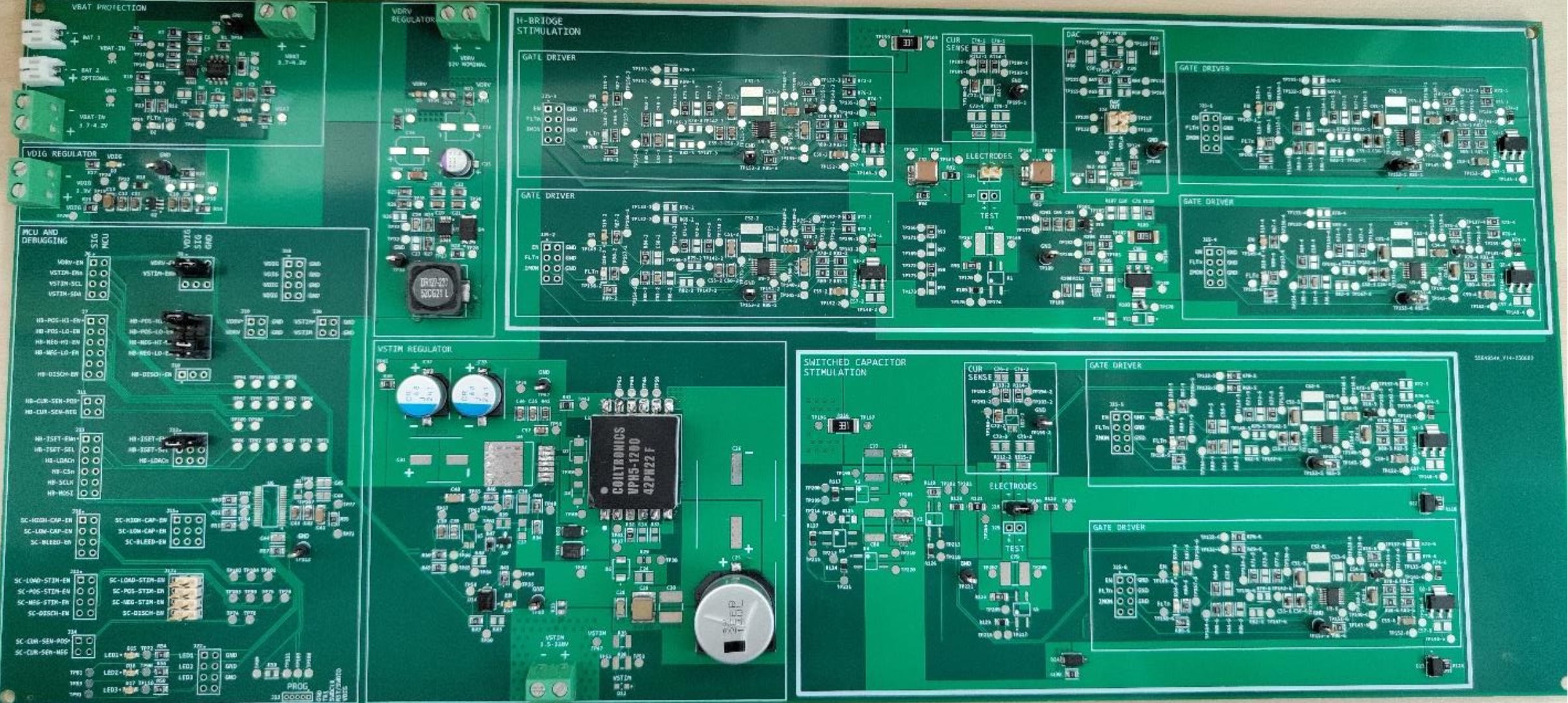}
    \caption{A photo of the assembled PCB. }
    \label{fig:PCB_photo}
\end{figure}

A large printed circuit board (PCB), measuring 38cm by 17cm, was developed to comprehensively evaluate the performance of the power management circuits, the switched capacitor (SC) circuits, and a conventional current-regulated H-bridge circuit~\cite{kang20230,ghovanloo2004compact}. The layout of this testing board is presented in Fig.~\ref{fig:PCB}. This PCB was designed without constraints on size, allowing ample space for the intentional inclusion of multiple test points and functional blocks to facilitate detailed evaluation and debugging. The board features several power planes and offers flexibility in control signal input: either a microcontroller can be directly assembled onto the PCB, or logic signals can be provided externally. This modular design ensures adaptability and ease of experimentation for various configurations and testing scenarios. Fig.~\ref{fig:PCB_photo} shows a photo of the fully assembled PCB. 

\begin{figure}[!ht]
    \centering
    \includegraphics[width=.8\linewidth]{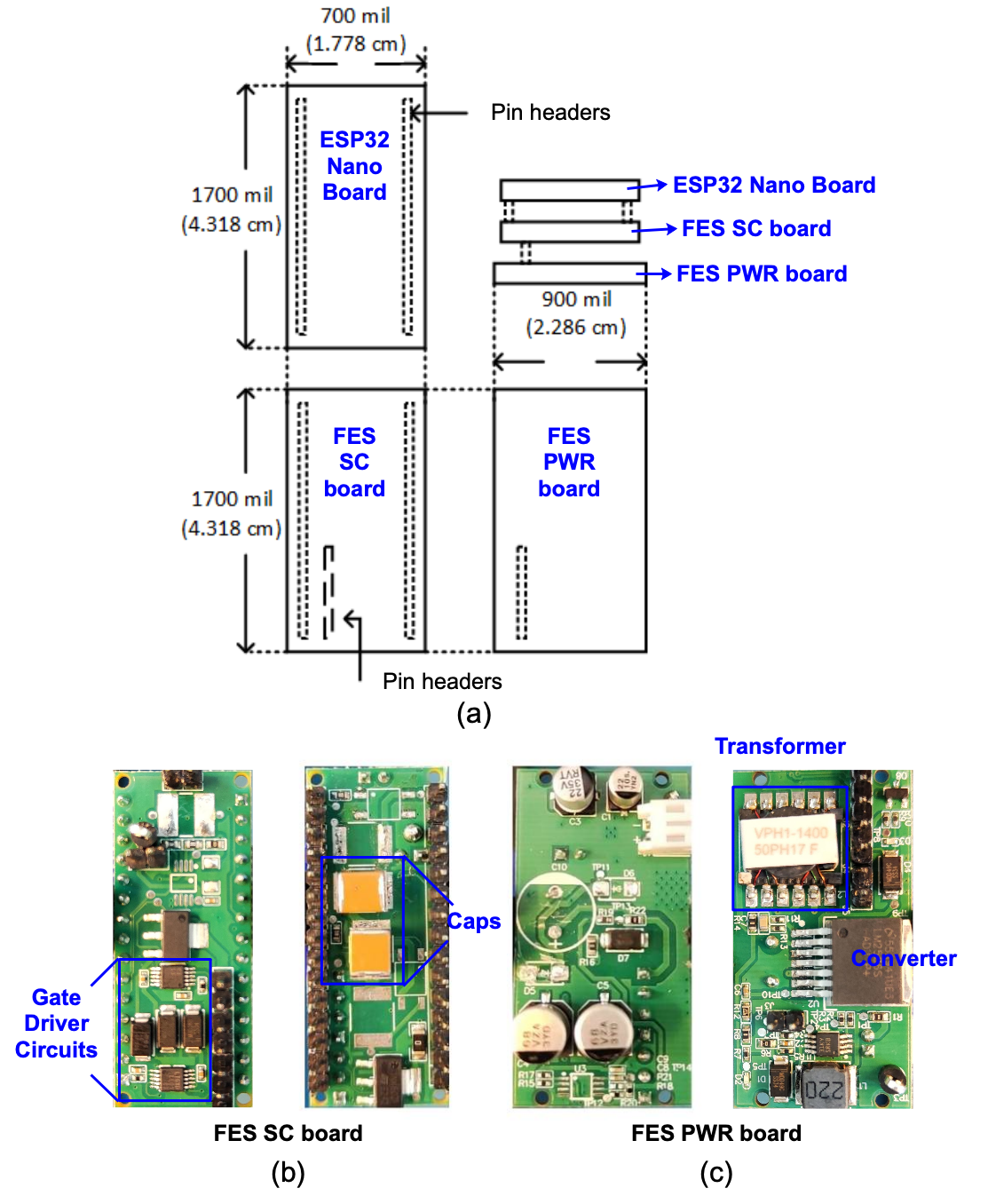}
    \caption{(a) An illustration of the PCB stacking design for the proposed HV stimulator. (b) Front and back photos of the board that consists of switched capacitor (SC) circuits. (b) Front and back photos of the board that consists of the power regulator and converter. }
    \label{fig:device}
\end{figure}

In addition, a compact HV FES device has been successfully prototyped and fully characterized. The assembled prototype measures 43mm $\times$ 30mm $\times$ 20mm, achieving a compact form factor suitable for wearable applications. To optimize the device footprint, the design incorporates a three-layer printed circuit board (PCB) structure, as depicted in Fig.~\ref{fig:device}~(a).

The top PCB features a commercially available Arduino Nano ESP32 microcontroller, which serves as the central control unit, facilitating parameter configuration and system timing~\cite{liu2014pennbmbi}. The middle PCB houses the core switched capacitor (SC) circuits, which are integral to the device’s ability to deliver precise stimulation pulses, as shown in Fig.~\ref{fig:device}(b). The bottom PCB contains the power regulation and conversion circuitry, including the flyback converter and associated components, which ensure the stable generation of high compliance voltages.

This layered PCB design not only minimizes the overall dimensions of the device but also enhances modularity, allowing for easier debugging, prototyping, and potential future upgrades. The compact design highlights the feasibility of integrating advanced FES functionality into a form factor that is both portable and practical for clinical and wearable applications.

The primary specifications of the developed high-voltage (HV) Functional Electrical Stimulation (FES) device are presented in Table~\ref{tab:spec}. The device offers a programmable output voltage that can be adjusted up to 135V, and an output current capable of reaching up to 20mA. Both parameters are configurable with a 7-bit resolution, enabling precise control to accommodate diverse therapeutic requirements.
\begin{table}[!ht]
\centering
\caption{Key Specifications of the developed FES}
\begin{tabular}{|c|c|}
\hline
\textbf{Parameters} & \textbf{Specification} \\ \hline
System Size         & 43 x 23 x 20 mm      \\ \hline
Topology            & Programmable monophasic / biphasic     \\ \hline
Output Voltage      & Programmable from 0 to 135~V, 7 bit resolution   \\ \hline
Output Current      & Programmable from 0 to 20~mA, 7 bit resolution    \\ \hline
Pulse Frequency     & Programmable from 1~Hz to 10k~Hz         \\ \hline
Pulse Rising Time   & under 20 ns                  \\ \hline
\end{tabular}\label{tab:spec}
\end{table}

\begin{figure}[!ht]
    \centering
    \includegraphics[width=.8\linewidth]{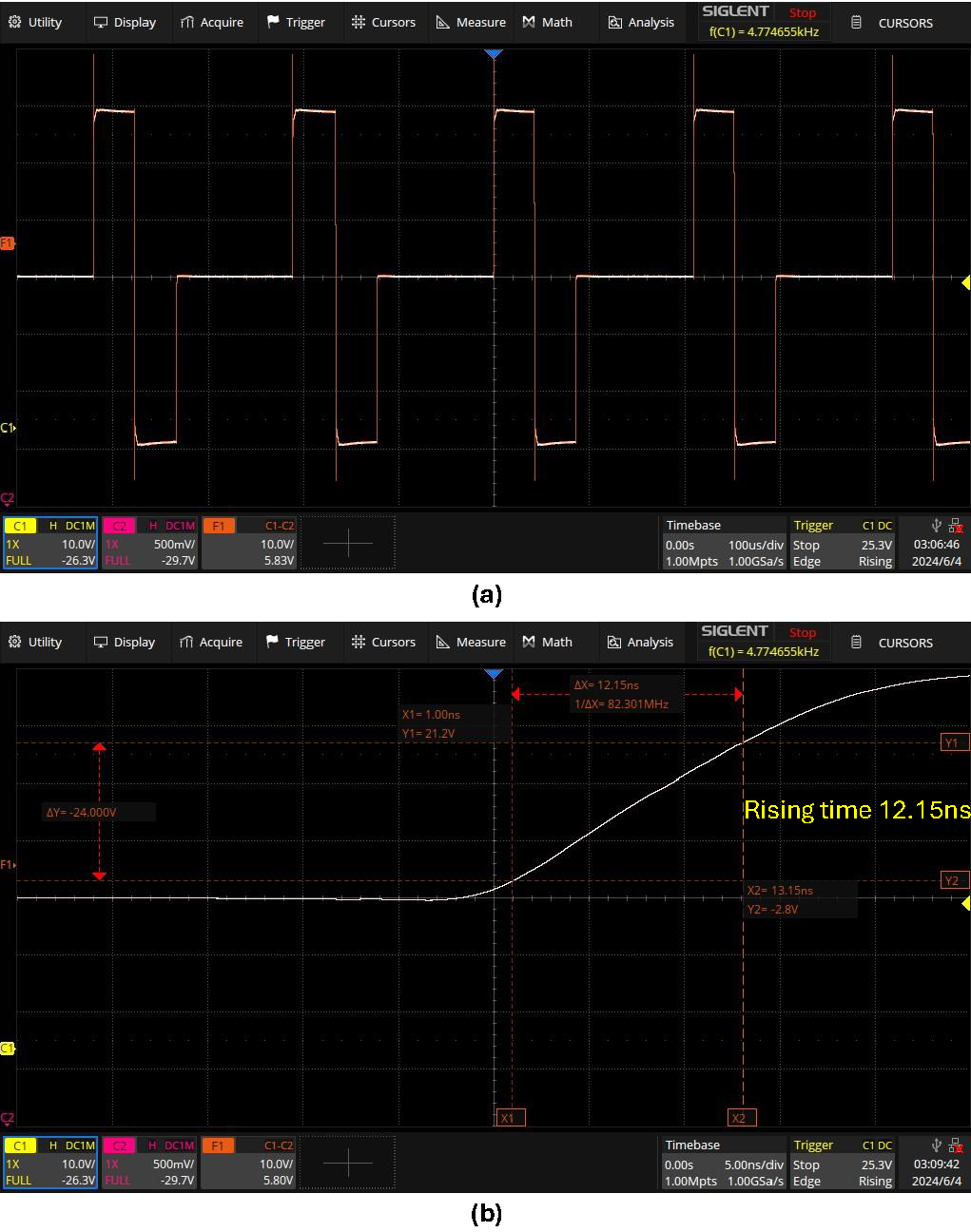}
    \caption{(a) Measured compliance voltage of the stimulator output during a biphasic stimulation. (b) A zoomed-in version of the rising edge measurement during the stimulation, showing a rise time of 12.15 ns for 90\% of the voltage swing.}
    \label{fig:exp}
\end{figure}

Figure~\ref{fig:exp}(a) illustrates the measured transient response of the stimulator during a biphasic stimulation cycle. A closer view of the rising edge is provided in Fig.\ref{fig:exp}(b), revealing a rise time of 12.15ns for 90\% of the voltage swing. These results validate the functionality of the device and confirm that its performance aligns with the established design targets. Before clinical trials, the prototypes will undergo comprehensive safety testing to ensure compliance with medical standards and patient safety requirements.

Table~\ref{tab:comparison} presents a comparison of the key specifications of the proposed system with those of state-of-the-art functional electrical stimulation (FES) systems reported in the literature and commercially available. In contrast to conventional designs, the proposed system achieves a combination of high output voltage and rapid rise times while maintaining a compact form factor and low power consumption.

\begin{table*}[!ht]
\centering
\caption{Comparison with prior high-voltage FES systems}
\label{tab:comparison}
\begin{tabular}{|c|c|c|c|c|c|}
\hline
\textbf{Publications}  & \textbf{Topology}   & \textbf{Rise Time}  & \textbf{Key Parameters} & \textbf{Size}    \\ \hline
Yigit et al. 2019\cite{Yiğit2019charge}                     & Biphasic             & N/A                  &   10 V, up to 1.4 mA        &  540 x 160 um chip    \\ \hline

Shirafkan et al. 2020\cite{Shirafkan2020}                   & Biphasic             & N/A                       &   80 V, up to 60 mA         &  96 x 89 mm  board     \\ \hline

Uehlin et al. 2020\cite{Uehlin2020}                         & Biphasic             & N/A                       &   11 V                      &   2 x 2 mm chip       \\ \hline

Palomeque-Mangut et al. 2022\cite{Mangut2022experimental}   & Biphasic             & N/A                 &   up to 12.5 V, 0-2.08 mA   & 1.3 x 1.8 mm chip      \\ \hline

Huang et al. 2023\cite{huang2023inductorless}               & Biphasic             & N/A                  & 40.5 V, up to 30 mA       & 1.61 x 1.68 mm chip     \\ \hline

Collu et al. 2023\cite{Collu2023}                           & Biphasic              & N/A                 &  90 V, up to 25 mA         & 115.9 x 61 mm board     \\ \hline

Danesh et al. 2024\cite{danesh2024cmos}                     & Monophasic/biphasic   & N/A                  &  
20 V, up to 14 mA         & 5.5 x 5.5 mm  chip   \\ \hline

This work                                                   & Monophasic/biphasic   & 20 ns                 & up to 135V, 0-20 mA       & 43 x 23 x 20 mm board \\ \hline
\end{tabular}
\end{table*}

\section{Conclusion}

This work presents the design and prototyping of an innovative high-voltage (HV) wearable stimulator for functional electrical stimulation (FES). By incorporating a switched capacitor (SC) circuit with current-sensing feedback, the device achieves rapid switching transitions, thereby minimizing user discomfort during therapy. High stimulation voltages are generated through an integrated flyback converter, enabling the device to operate efficiently on battery power. The system was constructed using low-cost, commercially available components, ensuring feasibility for reproduction in research and experimental applications. This design demonstrates significant potential to expand the therapeutic capabilities and applications of FES, making the technology more accessible and adaptable to a wider range of patient needs.

\bibliographystyle{IEEEtran}
\bibliography{ref}

\begin{thebibliography}{10}
\providecommand{\url}[1]{#1}
\csname url@samestyle\endcsname
\providecommand{\newblock}{\relax}
\providecommand{\bibinfo}[2]{#2}
\providecommand{\BIBentrySTDinterwordspacing}{\spaceskip=0pt\relax}
\providecommand{\BIBentryALTinterwordstretchfactor}{4}
\providecommand{\BIBentryALTinterwordspacing}{\spaceskip=\fontdimen2\font plus
\BIBentryALTinterwordstretchfactor\fontdimen3\font minus \fontdimen4\font\relax}
\providecommand{\BIBforeignlanguage}[2]{{%
\expandafter\ifx\csname l@#1\endcsname\relax
\typeout{** WARNING: IEEEtran.bst: No hyphenation pattern has been}%
\typeout{** loaded for the language `#1'. Using the pattern for}%
\typeout{** the default language instead.}%
\else
\language=\csname l@#1\endcsname
\fi
#2}}
\providecommand{\BIBdecl}{\relax}
\BIBdecl

\bibitem{popovic2014advances}
D.~B. Popovi{\'c}, ``Advances in functional electrical stimulation (fes),'' \emph{Journal of Electromyography and Kinesiology}, vol.~24, no.~6, pp. 795--802, 2014.

\bibitem{peckham2005functional}
P.~H. Peckham and J.~S. Knutson, ``Functional electrical stimulation for neuromuscular applications,'' \emph{Annu. Rev. Biomed. Eng.}, vol.~7, pp. 327--360, 2005.

\bibitem{lynch2008functional}
C.~L. Lynch and M.~R. Popovic, ``Functional electrical stimulation,'' \emph{IEEE control systems magazine}, vol.~28, no.~2, pp. 40--50, 2008.

\bibitem{Huerta2012}
S.~C. Huerta, M.~Tarulli, A.~Prodic, M.~R. Popovic, and P.~W. Lehn, ``A universal functional electrical stimulator based on merged flyback-sc circuit,'' in \emph{2012 15th International Power Electronics and Motion Control Conference (EPE/PEMC)}, 2012, pp. LS5a.3--1--LS5a.3--5.

\bibitem{van2016system}
J.~Van~der Spiegel, M.~Zhang, and X.~Liu, ``System-on-a-chip brain-machine-interface design-a review and perspective,'' in \emph{2016 13th IEEE International Conference on Solid-State and Integrated Circuit Technology (ICSICT)}.\hskip 1em plus 0.5em minus 0.4em\relax IEEE, 2016, pp. 203--206.

\bibitem{vidal2010towards}
J.~Vidal and M.~Ghovanloo, ``Towards a switched-capacitor based stimulator for efficient deep-brain stimulation,'' in \emph{2010 Annual International Conference of the IEEE Engineering in Medicine and Biology}.\hskip 1em plus 0.5em minus 0.4em\relax IEEE, 2010, pp. 2927--2930.

\bibitem{liu2020fully}
X.~Liu, H.~Zhu, T.~Qiu, S.~Y. Sritharan, D.~Ge, S.~Yang, M.~Zhang, A.~G. Richardson, T.~H. Lucas, N.~Engheta \emph{et~al.}, ``A fully integrated sensor-brain--machine interface system for restoring somatosensation,'' \emph{IEEE Sensors Journal}, vol.~21, no.~4, pp. 4764--4775, 2020.

\bibitem{blade2023semg}
S.~Blade, Z.~Yao, Y.~Hou, Y.~Li, S.~Zhou, Y.~Wang, and X.~Liu, ``An semg-controlled prosthetic hand featuring a tiny cnn-transformer model and force feedback,'' in \emph{2023 IEEE Biomedical Circuits and Systems Conference (BioCAS)}.\hskip 1em plus 0.5em minus 0.4em\relax IEEE, 2023, pp. 1--5.

\bibitem{liu2015pennbmbi}
X.~Liu, M.~Zhang, B.~Subei, A.~G. Richardson, T.~H. Lucas, and J.~Van~der Spiegel, ``The pennbmbi: Design of a general purpose wireless brain-machine-brain interface system,'' \emph{IEEE transactions on biomedical circuits and systems}, vol.~9, no.~2, pp. 248--258, 2015.

\bibitem{ibitoye2016strategies}
M.~O. Ibitoye, N.~A. Hamzaid, N.~Hasnan, A.~K. Abdul~Wahab, and G.~M. Davis, ``Strategies for rapid muscle fatigue reduction during fes exercise in individuals with spinal cord injury: a systematic review,'' \emph{PloS one}, vol.~11, no.~2, p. e0149024, 2016.

\bibitem{kang20230}
V.~Kang, A.~A. Malak, K.~Lau, Y.~Zhu, C.~Morshead, and X.~Liu, ``A 0.9 g battery-free wireless stimulator with infrared communication for applications in neural repair and regeneration,'' in \emph{2023 IEEE International Symposium on Circuits and Systems (ISCAS)}.\hskip 1em plus 0.5em minus 0.4em\relax IEEE, 2023, pp. 1--5.

\bibitem{ghovanloo2004compact}
M.~Ghovanloo and K.~Najafi, ``A compact large voltage-compliance high output-impedance programmable current source for implantable microstimulators,'' \emph{IEEE Transactions on Biomedical Engineering}, vol.~52, no.~1, pp. 97--105, 2004.

\bibitem{liu2014pennbmbi}
X.~Liu, B.~Subei, M.~Zhang, A.~G. Richardson, T.~H. Lucas, and J.~Van~der Spiegel, ``The pennbmbi: A general purpose wireless brain-machine-brain interface system for unrestrained animals,'' in \emph{2014 IEEE International Symposium on Circuits and Systems (ISCAS)}.\hskip 1em plus 0.5em minus 0.4em\relax IEEE, 2014, pp. 650--653.

\bibitem{Yiğit2019charge}
H.~A. Yiğit, H.~Uluşan, S.~Chamanian, and H.~Külah, ``Charge balance circuit for constant current neural stimulation with less than 8 nc residual charge,'' in \emph{2019 IEEE International Symposium on Circuits and Systems (ISCAS)}, 2019, pp. 1--5.

\bibitem{Shirafkan2020}
\BIBentryALTinterwordspacing
R.~Shirafkan, O.~Shoaei, and M.~K. Ahmadi, ``A high efficient adiabatic transcutaneous electrical nerve stimulator (tens) with current regulation,'' \emph{AEU - International Journal of Electronics and Communications}, vol. 123, p. 153275, 2020. [Online]. Available: \url{https://www.sciencedirect.com/science/article/pii/S1434841120307366}
\BIBentrySTDinterwordspacing

\bibitem{Uehlin2020}
J.~P. Uehlin, W.~A. Smith, V.~R. Pamula, E.~P. Pepin, S.~Perlmutter, V.~Sathe, and J.~C. Rudell, ``A single-chip bidirectional neural interface with high-voltage stimulation and adaptive artifact cancellation in standard cmos,'' \emph{IEEE Journal of Solid-State Circuits}, vol.~55, no.~7, pp. 1749--1761, 2020.

\bibitem{Mangut2022experimental}
D.~Palomeque-Mangut, {\'A}.~Rodr{\'\i}guez-V{\'a}zquez, and M.~Delgado-Restituto, ``Experimental validation of a high-voltage compliant neural stimulator implemented in a standard 1.8 v/3.3 v cmos process,'' in \emph{2022 IEEE Biomedical Circuits and Systems Conference (BioCAS)}.\hskip 1em plus 0.5em minus 0.4em\relax IEEE, 2022, pp. 335--339.

\bibitem{huang2023inductorless}
Y.-K. Huang and S.~Rodriguez, ``An inductorless 40.5 v high-voltage generator for integrated neuromuscular electrical stimulators,'' in \emph{2023 IEEE Biomedical Circuits and Systems Conference (BioCAS)}.\hskip 1em plus 0.5em minus 0.4em\relax IEEE, 2023, pp. 1--5.

\bibitem{Collu2023}
\BIBentryALTinterwordspacing
R.~Collu, R.~Paolini, M.~Bilotta, A.~Demofonti, F.~Cordella, L.~Zollo, and M.~Barbaro, ``Wearable high voltage compliant current stimulator for restoring sensory feedback,'' \emph{Micromachines}, vol.~14, no.~4, 2023. [Online]. Available: \url{https://www.mdpi.com/2072-666X/14/4/782}
\BIBentrySTDinterwordspacing

\bibitem{danesh2024cmos}
A.~R. Danesh, H.~Pu, M.~Safiallah, A.~H. Do, Z.~Nenadic, and P.~Heydari, ``A cmos bd-bci: Neural recorder with two-step time-domain quantizer and multi-polar stimulator with dual-mode charge balancing,'' \emph{IEEE Transactions on Biomedical Circuits and Systems}, 2024.

\end{thebibliography}

\ifCLASSOPTIONcaptionsoff
  \newpage
\fi

\end{document}